\documentclass[twocolumn,prl,epsfig,amsmath,amssymb]{revtex4}
\usepackage{graphicx}

\begin{document}

\title{Does Probability become Fuzzy in Small Regions of Spacetime?}

\author{Markus M\"uller\footnote{E-mail: mueller@math.tu-berlin.de}\\}
\mbox{}\vskip 0.5cm \affiliation{
$^1$Institut f\"ur Mathematik, Technische Universit\"at Berlin, Stra\ss e des 17.~Juni 136, 10623 Berlin, Germany
\\
$^2$Max Planck Institute for Mathematics in the Sciences, Inselstr.~22, 04103 Leipzig, Germany\\
}
\date{February 3, 2009}
\begin{abstract}
In a recent paper, Buniy et al. have argued that a possible discretization of spacetime leads to an
unavoidable discretization of the state space of quantum mechanics. In this paper, we show that this
conclusion is not limited to quantum theory: in any classical, quantum, or more general probabilistic theory,
states (i.e. probabilities or corresponding amplitudes) become discrete or fuzzy for observers, as long as time evolution is reversible
and entropy is locally bounded.

Specifically, we show that the Bekenstein bound suggests that probabilities in small
closed regions of space carry an uncertainty inversely proportional to the square root of the system's effective radius and energy.
\end{abstract}


\maketitle

Is Hilbert Space discrete? A recent paper by Buniy et al.~\cite{Buniy1} with this title claims that
a possible discretization of spacetime leads to a discretization of the state space of quantum mechanics.
We briefly review their arguments as follows: given some unknown quantum state, say, a spin-$\frac 1 2$ particle (a qubit)
\[
   |\psi\rangle=\cos\theta |+\rangle+e^{i\phi}\sin\theta |-\rangle,
\]
the phases $\theta$ and $\phi$ can be determined by state tomography, using appropriate measuring apparatuses. But if spacetime is discrete, there
is only a finite number of measuring apparatuses that can possibly be constructed. This means that measurements
can only be done within some fixed accuracy which is determined by the coarseness of spacetime.
In particular, interpreting $\theta$ and $\phi$ as angular variables,
there are only finitely many possible rotations of a measuring apparatus, such that those variables cannot be
measured with greater accuracy than about $1/d$, where $d$ is the size of the apparatus.

For the details of the argumentation, we refer the reader to~\cite{Buniy1} and a related paper~\cite{Buniy2}.
In this paper, we give a more elementary and rigorous derivation, which shows that this result has more
general validity and is not restricted to quantum theory.

We start with a simple thought experiment. Suppose an observer is confined to a small region of space,
say, an astronaut is captured inside a free-falling spaceship (or elevator, as is frequently used
in thought experiments on the equivalence principle). The observer is completely isolated from the outside world,
that is, we assume that the spaceship is a perfectly closed system.

Moreover, suppose that the astronaut is given some random experiment with two different outcomes (``success''
and ``failure'') that he can in principle repeat an arbitrary number of times. The probability of success is given by a
real number $p^*\in[0,1]$ that the astronaut does not know. The astronaut's task is to determine the unknown
probability value $p^*$. For example,
\begin{itemize}
\item the astronaut might possess some coin which is loaded instead of fair, and he wants to know
the probability of ``heads'' (classical probability), or
\item the astronaut carries a single spin-$\frac 1 2$ particle that is initialized by some black box
in an unknown quantum state $|\psi\rangle$, and he wants to find out the probability of ``spin up''
(quantum state).
\end{itemize}
The only way for the observer to estimate $p^*$ is
to repeat the random experiment a large number of times,
remember the number $k$ of ``success'' outcomes and the number $m$ of total trials, and use
$k/m$ as an estimator of $p^*$.

This way, the astronaut can determine $p^*$ in principle to arbitrary accuracy by choosing $m$ large enough,
that is, by repeating the random experiment often enough.
But there is a small detail that may
cause problems: every trial produces one bit of information, say, ``0''
for ``failure'' or ``1'' for ``success''. After $m$ trials, there are $m$ useless bits that have
accumulated inside the spaceship, keeping a record of the sequence of measurement outcomes.

Those bits cannot simply be disposed to the environment, because the spaceship is by assumption completely isolated.
Also, since time evolution is reversible (which is true for classical as well as for quantum mechanics), the bits
cannot be directly erased, but they can only be converted into different forms of information inside the spaceship. For example,
if the astronaut saves the measurement results on a computer hard disc and deletes the corresponding file after a million
trials, global reversibility ensures that the million bits are simply transmitted to different degrees of freedom inside the
spaceship, for example to the vibrational modes of the air's molecules. The same is true for the initialization of the random experiment.


As a consequence, $m$ bits of information are accumulated inside the spaceship if the random experiment is performed $m$
times. For simplicity of the argument, let's assume that the value of $p^*$ is not too far away from $1/2$, such that
the outcome bit string of length $m$ cannot be significantly compressed.
If for some reason the number of bits that the spaceship can hold is bounded
(e.g. by the laws of physics), then the astronaut is forced to stop performing trials at a certain point --- namely,
at the point when his spaceship ``runs out of memory''.

Note that this kind of argumentation is similar to the resolution of the thought experiment of Maxwell's demon~\cite{Maruyama},
where the fact that the demon runs out of memory and has to erase information (which is costly) preserves the
second law of thermodynamics.

In fact, Bekenstein~\cite{Bekenstein}
has derived an upper bound on the number of bits $I$ that can be stored inside a bounded spatial region
of effective radius $R$ and energy $E$, based on calculations in quantum field theory and black hole entropy. It is given by
\begin{equation}
   I\leq\frac{2\pi E R}{\hbar c \ln 2}.
   \label{EqBekenstein}
\end{equation}
This means that the actual laws of physics, as we know them, indeed force the astronaut to stop
performing the random experiment after a large, but finite number of trials. Hence the accuracy
to determine $p^*$ by measurements is bounded. We will now compute the resulting unavoidable uncertainty
by using elementary probability theory.

To do this, we treat $p^*$ as a random variable, assuming uniform prior distribution on the interval $[0,1]$.
In case that we repeat a random experiment $m$ times and have $k$ successes (and $m-k$ failures), it is well-known
\cite[p. 165]{Jaynes} that the probability density of $p^*$ is then given by
\[
   {\rm Prob}(p^*=\theta\,|\,k)=
   \frac{(m+1)!}{k!(m-k)!}\theta^k(1-\theta)^{m-k}=:f(\theta).
\]
Defining the uncertainty of measuring $p^*$ in the usual way as $\Delta p^*:=\sqrt{{\rm Var}(p^*)}$,
we get
\begin{eqnarray*}
(\Delta p^*)^2&=&{\rm Var}(p^*)=E(p^{*2})-(E(p^*))^2\\
&=&\int_0^1 \theta^2 f(\theta)d\theta-\left(\int_0^1 \theta \,\,f(\theta)d\theta\right)^2\\
&=&\frac{(k+1)(m-k+1)}{(m+2)^2(m+3)}\approx \frac 1 {4m}
\end{eqnarray*}
for large $m\in\mathbb{N}$ and $k\approx m/2$. (A rigorous worst-case lower bound for arbitrary $k$ would
be $(\Delta p^*)^2\geq \frac 1 {18 m^2}$, but we have assumed previously that $p^*$ does not deviate too much from $1/2$,
such that $k\approx m/2$ holds.)

In conclusion, we get $\Delta p^*\approx 1/(2\sqrt{m})$. That is, {\em if the amount of information that the spaceship can
hold is upper bounded by some maximum value of $m$ bits, then the astronaut cannot determine unknown probability values with
greater accuracy than about $1/(2\sqrt{m})$.}

If we consider the whole universe as our ``spaceship'' and use the estimate~\cite{Lloyd} of about $10^{90}$ as the number of bits in our universe,
we get the approximate value $\Delta p^*\approx 10^{-45}$, which means that probabilities so far could not have been determined
by any observer in our universe's history to more than about $45$ digits of precision.

In particular, if we assume the laws of physics as we know them, we can use Eq.~(\ref{EqBekenstein}) and obtain,
omitting all multiplicative constants of order one,
\[
   \Delta p^* \gtrsim \sqrt{\frac{\hbar c}{E R}}.
\]
This inequality gives an approximate lower bound on the unavoidable uncertainty of probability measurements in closed spatial systems of
effective energy $E$ and radius $R$, in the case that $p^*\approx 1/2$.

This result is comparable to that by Buniy et al.~\cite{Buniy1}, but we have not used any specifically quantum property
in our analysis. In fact, we have derived the uncertainty bound $\Delta p^*\approx \frac 1 {2\sqrt m}$ under the following
two assumptions only:
\begin{itemize}
\item[1.] Entropy is locally bounded (by some number $m$),
\item[2.] time evolution is reversible.
\end{itemize}
Thus, the bound is valid not only for quantum mechanics, but also for classical statistical mechanics
or even more exotic probabilistic theories that generalize quantum theory (such theories have recently attracted
attention, cf.~\cite{Barrett} and the references therein).

Our result says that the possibility of observers to determine the value of probabilities is limited.
Does this mean that probability itself is indeed ``fuzzy'' or even ``discrete'' as proposed in the quantum
setting by Buniy et al.~\cite{Buniy1}?

We leave this question for others to speculate and give only one concluding remark. The proposal in~\cite{Buniy1}
to modify the state space of a qubit, i.e.
the Bloch sphere, looks somewhat similar to the considerations on general probabilistic theories in~\cite{Barrett}:
such theories can have convex state spaces for two-level systems that are ``more general'' than the Bloch sphere.
If the state space was modified when the region of space becomes small, then some previously pure states
might become mixed. It might be interesting to speculate if this mechanism can be an interpretation of ``probability
getting fuzzy'' in small regions of space.

\vskip 0.25cm \noindent{\bf Acknowledgments}
I am grateful to Greg ver Steeg and Arleta Szko\l a for helpful and interesting discussions.

\end{document}